\documentclass[twoside,slac_one]{revtex4}
\usepackage{graphicx}
\usepackage{fancyhdr}
\usepackage{amsmath} % American Mathematics Society standards
\usepackage{bm}% bold math
\usepackage{amsxtra}
\usepackage{amssymb}
\usepackage{amsthm}
\usepackage{latexsym}
\usepackage{lscape}

\begin{document}

%Title of paper
\title{The ArgoNeuT experiment}

% Repeat the \author .. \affiliation  etc. as needed
%
% \affiliation command applies to all authors since the last
% \affiliation command. The \affiliation command should follow the
% other information

\author{R. Gu\'enette}
\affiliation{Department of Physics, Yale University, New Haven, CT, USA}

\begin{abstract}
The ArgoNeuT experiment features a 175 liter Liquid Argon (LAr) Time Projection Chamber (TPC) that was located upstream of the MINOS near detector in the NuMI neutrino beam at Fermilab. The project is part of the LAr TPC development program in the US and has helped initiate the development of simulation and reconstruction tools for LAr TPCs. In addition to its development goals, ArgoNeuT will perform several cross-section measurements on Ar in the few-GeV energy range. A total of 1.35E20 Protons on Target were accumulated and data analysis is ongoing. I will review the experiment and its status, as well as preliminary results from the data analysis.
\end{abstract}

%\maketitle must follow title, authors, abstract
\maketitle

\thispagestyle{fancy}

% body of paper here - Use proper section commands
% References should be done using the \cite, \ref, and \label commands
% Put \label in argument of \section for cross-referencing
%\section{\label{}}

%%%%%%%%%%%%%%%%%%%%%%%%%%%%%%%%%%
\section{Introduction}

Liquid Argon (LAr) Time Projection Chambers (TPCs) seem to be a very promising technology for the next generation of neutrino detectors. This next generation will require multi-kilo-ton detectors, and LAr TPCs are an attractive option. These detectors offer high detection efficiencies and excellent background rejection and they appear to be scalable. However, the scalability has not yet been fully demonstrated. The ArgoNeuT experiment, a $175l$ LAr TPC was a first step to a $R\&D$ program in the United States. The main goal of ArgoNeuT was to gain experience building and operating LAr TPCs. In addition, the experiment collected data in an energy range relevant to neutrino oscillation physics. These data were used to develop simulation and analysis tools and the first physics results extracted from the data were recently presented. A review of the ArgoNeuT experiment will be given as well as a description of the data analysis and an overview of the first results on $\nu_{\mu}$ Charged-Current (CC) inclusive differential cross-section measurement by ArgoNeuT.

%%%%%%%%%%%%%%%%%%%%%%%%%%%%%%%%%%
\section{The ArgoNeuT detector}

\begin{figure}[ht]
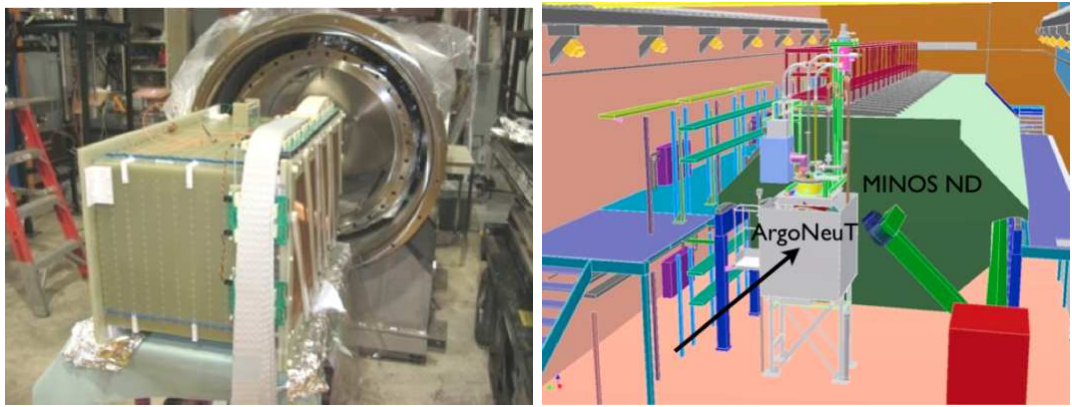

\centering
\begin{tabular}{cc}
\includegraphics[width=7cm]{tpc.epsi} &
\includegraphics[width=7cm]{det.epsi}
\end{tabular}
\caption{Left: The ArgoNeuT TPC about to enter the cryostat. Right: Schematics of the location of the ArgoNeuT detector in the NuMI beam, in front of the MINOS Near Detector (ND). The black arrow indicates the NuMI beam direction.} 
\label{fig1}
\end{figure}

The ArgoNeuT TPC is a 90cm $\times$ 40cm $\times$ 48cm rectangular volume containing $175l$ of LAr active volume. The TPC was positioned in a vacuum jacketed cryostat. Left picture in Figure \ref{fig1} shows the TPC about to enter the cryostat. The TPC consists of three wire planes with 240 wires each, separated by 4mm. The innermost wire plane has vertical wire orientation and was not instrumented. The second plane, the instrumented induction plane, has wires oriented at $+60^{\circ}$ and the third plane, the collection plane, has wires oriented at $-60^{\circ}$. The wire planes can be seen in the left panel of Figure \ref{fig1} on the right of the TPC. A 25kV voltage was applied to the cathode, creating a 500V/cm electric field. More details on the detector, its electronics and the cryogenics can be found in \cite{mitch2009}.\\

The ArgoNeuT detector was placed in the NuMI tunnel at Fermilab, just in front of the MINOS near detector \cite{minos}. Since ArgoNeuT's TPC is too small to contain the majority of the muons produced in neutrino interactions from the energetic NuMI beam, the information from the MINOS near detector will be used in the ArgoNeuT data analysis. This is an enormous advantage for ArgoNeuT since MINOS, in addition to provide information on the momentum of the muons, can also determine the sign of the muon from its magnetized detector.\\

\begin{figure}[ht]
\centering
\includegraphics[width=11cm]{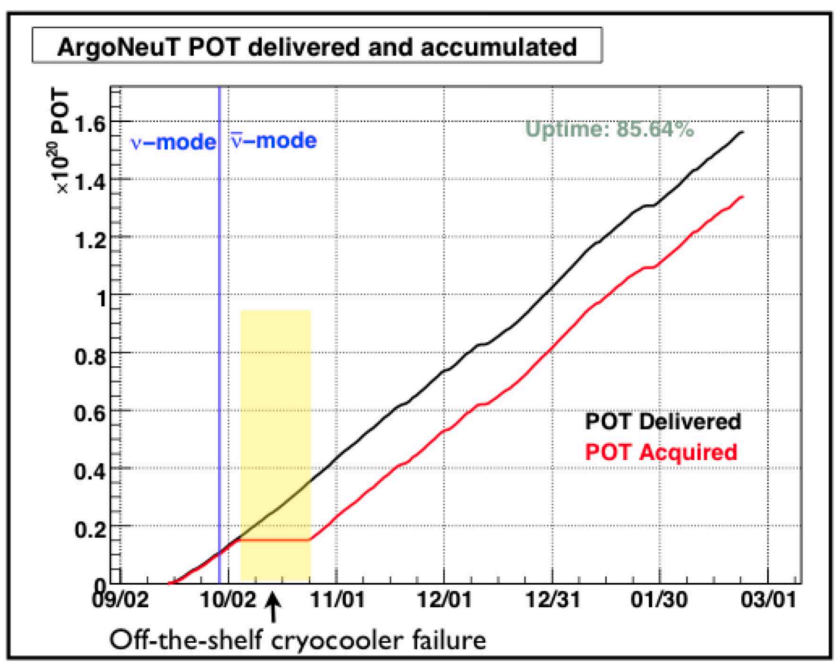}
\caption{Amount of POTs delivered to and accumulated by ArgoNeut during the $\sim$5 months of data taking. The yellow stripe shows the time where the detector was down because of the cryocooler failure.} 
\label{pot}
\end{figure}

The ArgoNeuT detector took data between September 2009 and February 2010. Data taking was very stable during over five months, only $\sim$ two weeks were lost because of an off-the-shelf cryo-cooler failure. A total of two weeks of data were acquired in neutrino mode, corresponding to $8.5\times10^{18} POTs$ and the rest of the time was taken in anti-neutrino mode, for a total of $1.25\times10^{20} POTs$. Figure \ref{pot} shows the amount of POTs acquired by ArgoNeuT in comparison to the amount of POTs delivered. \\

%%%%%%%%%%%%%%%%%%%%%%%%%%%%%%%%%%
\section{Data analysis}

The ArgoNeuT collaboration uses an analysis software designed for all LAr experiments in the US. This software, LArSoft \cite{larsoft}, was heavily developed by the ArgoNeuT collaboration, since it is the first experiment to have data. This software allows to reconstruct events from the raw signal on the individual wires.\\

The reconstruction chains starts with a hit construction and identification from the raw data signal information. Once the hits are identified, they are clustered with other nearby hits. Two dimensional lines are then reconstructed to identify the line-like tracks. Finally, a three dimensional track reconstruction is performed. More details on the reconstruction can be found in \cite{josh}.\\

As mentioned previously, ArgoNeuT uses the data from the MINOS detector to obtain the information on muons that escape the detector. At this step, identified muon tracks in ArgoNeuT are matched to muon tracks in MINOS. The details in the matching requirements can be found here \cite{josh}. Once the matching is successful, the charge and momentum of the muons are extracted from MINOS data.\\

%%%%%%%%%%%%%%%%%%%%%%%%%%%%%%%%%%
\section{Cross-section measurement}

\begin{figure}[ht]
\centering
\includegraphics[angle=-90,width=11cm]{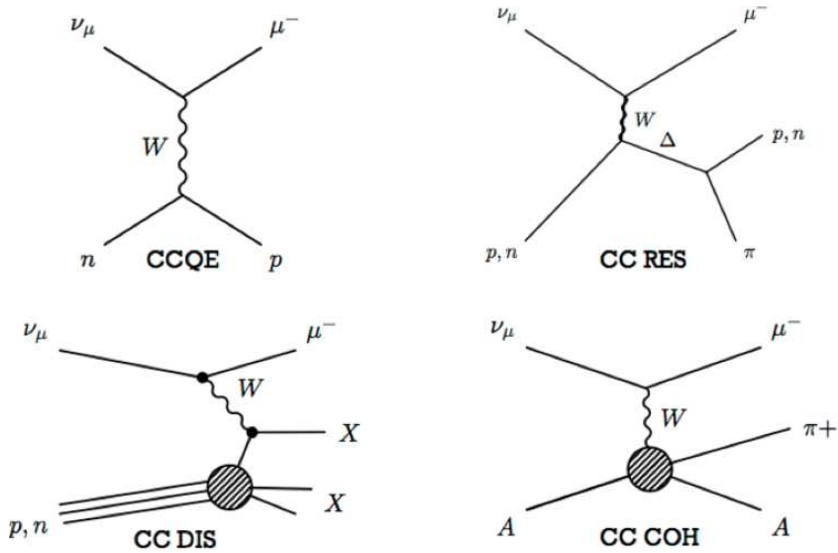}
\caption{Different $\nu_{\mu}$ CC interactions. The top left diagram shows the charged-current quasi-elastic (CCQE) interaction, the top right shows the charged-current resonant (CC RES) interaction, the bottom left shows the charged-current deep inelastic scattering (CC DIS) interaction and the bottom right shows the charged-current coherent (CC COH) interaction.} 
\label{interactions}
\end{figure}

An important goal of the ArgoNeuT experiment is to measure neutrino cross sections on Ar. The first natural measurement to be done with the data is an all-inclusive charge-current measurement. Indeed, such a measurement is totally independent on channel definitions and is minimally sensitive to final-state interactions (FSIs). As seen in Figure \ref{interactions}, all the different types of CC interactions have in common an outgoing muon. In addition, further cross-section measurements can be compared to the CC-inclusive one to help disentangling the effects of FSIs and nuclear modeling. In order to provide kinematic information, the differential cross section, in function of the outgoing muon angle and momentum is a very useful measurement for theoretical predictions. Normally, the double differential cross-section ($\frac{d\sigma}{d\theta_{\mu}dp_{\mu}}$) would be calculated, but in order to do so, high statistics are required. Since ArgoNeuT only collected two weeks of neutrino data, only one dimensional differential cross sections ($\frac{d\sigma}{d\theta_{\mu}}, \frac{d\sigma}{dp_{\mu}}$) are calculated.\\

%\begin{equation}
%\frac{d\sigma}{d\theta_{\mu}},
%\frac{d\sigma}{dp_{\mu}}\\
%%\caption{One dimensional differential cross sections in function of the muon kinematics (angle and momentum).}
%\label{equ2}
%\end{equation}
%\begin{equation}
%\frac{d\sigma}{d\theta_{\mu}dp_{\mu}}
%\caption{Two dimensional differential cross section {One dimensional differential cross sections in function of the muon kinematics (angle and momentum).}
%\label{equ1}
%\end{equation}

%\vspace{1cm}

The total CC cross section as a function of neutrino energy is also a useful measurements. However, due the modest size of the ArgoNeuT TPC, events are rarely fully reconstructed and it is therefore not possible to reconstruct the total energy of all events.\\

\begin{figure}[ht]
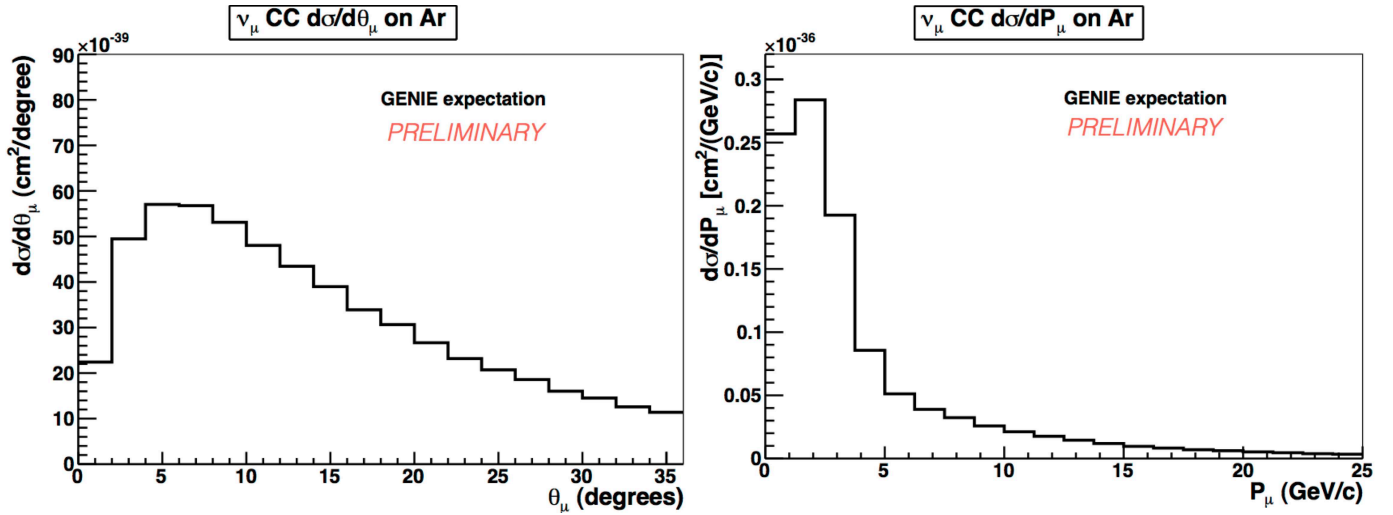

\centering
\begin{tabular}{cc}
\includegraphics[angle=-90,width=9cm]{thetaArdiff_MC.epsi}&
\includegraphics[angle=-90,width=9cm]{momArdiff_MC.epsi}
\end{tabular}
\caption{Left: MC expectations for the CC $\nu_{\mu}$ differential cross section in muon angle on an argon target for ArgoNeuT. Right: MC expectations for the CC $\nu_{\mu}$ differential cross section in muon momentum on an argon target for ArgoNeuT.  The differential cross sections are reported per ``argon nucleon''. Figures from \cite{josh}} 
\label{xsec}
\end{figure}

All the details of the analysis and results of the CC-inclusive differential cross-section measurements can be found here \cite{josh} and will be available in a paper in preparation. In this proceeding, only the differential cross-section MC expectations are presented. Figure \ref{xsec} shows the CC-inclusive differential cross-section predicted from simulations in function of the muon angle (left) and of the muon momentum (right).

\section{Conclusions}

ArgoNeuT ran stably for over five months, collecting an impressive sample of neutrino and anti-neutrino data. The two weeks of neutrino data has started to be analyzed and the simulations predictions for the $\nu_{\mu}$ CC-inclusive differential cross section were reported here. More results will be presented in a paper in preparation. Future analysis will allow channel-specific cross-section measurements. The sample of anti-neutrino data, containing about fifteen times more statistics, will also be analyzed in the near future.\\ 

This measurement also demonstrates that the LArSoft software is operational and ready for further analyses. The ArgoNeuT data will also be used to look in details at FSIs to study nuclear effects.\\

%%%%%%%%%%%%%%%%%%%%%%%%%%%%%%%%%%
\begin{acknowledgments}
The ArgoNeuT collaboration gratefully acknowledge the cooperation of the MINOS collaboration in providing their data for use in the analysis.
\end{acknowledgments}

\bigskip % extra skip inserted
% Create the reference section using BibTeX:
%\bibliography{basename of .bib file}

\end{document}